\documentclass[prl,aps,twocolumn,floats,nofootinbib]{revtex4-1}
\usepackage{amsmath,amssymb,graphicx,psfrag}
\usepackage[colorlinks,linkcolor={blue},citecolor={blue},urlcolor={blue}]{hyperref}
\usepackage{caption, subcaption}
\begin{document}
\renewcommand{\ni}{{\noindent}}
\newcommand{\dprime}{{\prime\prime}}
\newcommand{\be}{\begin{equation}}
\newcommand{\ee}{\end{equation}}
\newcommand{\bea}{\begin{eqnarray}} 
\newcommand{\eea}{\end{eqnarray}}
\newcommand{\la}{\langle}
\newcommand{\ra}{\rangle} 
\newcommand{\dg}{\dagger}
\newcommand\lbs{\left[}
\newcommand\rbs{\right]}
\newcommand\lbr{\left(}
\newcommand\rbr{\right)}
\newcommand\f{\frac}
\newcommand\e{\epsilon}
\newcommand\ua{\uparrow}
\newcommand\da{\downarrow}
\newcommand\mbf{\mathbf}

\newcommand{\Eqn}[1] {Eqn.~(\ref{#1})}
\newcommand{\Fig}[1]{Fig.~\ref{#1}}

\title{Local density of states and scattering rates across the many-body localization transition}
\author{Atanu Jana$^{1}$, V. Ravi Chandra$^{1}$, Arti Garg$^{2}$} 
\affiliation{$^{1}$ School of Physical Sciences, National Institute of Science Education and Research Bhubaneswar, HBNI, Jatni, Odisha 752050, India}
\affiliation{$^{2}$ Theory Division, Saha Institute of Nuclear Physics, 1/AF Bidhannagar, Kolkata 700 064, India}
\vspace{0.2cm}
\begin{abstract}
  \vspace{0.3cm}

Characterizing the many-body localization (MBL) transition in strongly disordered and interacting quantum systems is an important issue in the field of condensed matter physics.
We study the single particle Green's functions for a disordered interacting system in one dimension using exact diagnonalization in the infinite temperature limit and provide strong evidence that single particle excitations carry signatures of delocalization to MBL transition. In the delocalized phase, the typical values of the local density of states and the scattering rate are finite while in the MBL phase, the typical values for both the quantities become vanishingly small.
The probability distribution functions of the local density of states and the scattering rate are broad log-normal distributions in the delocalized phase while the distributions become very narrow and sharply peaked close to zero in the MBL phase. We also study the eigenstate Green's function for all the many-body eigenstates and demonstrate that both, the energy resolved typical scattering rate and the typical local density of states, can track the many-body mobility edges.

\vspace{0.cm}
\end{abstract} 
\maketitle

The physics of Anderson localization~\cite{Anderson} in non-interacting disordered quantum systems has been a cornerstone of condensed matter theory. 
Turning on interactions in these disordered systems results in the many-body localized (MBL) phase where the system lacks 
transport~\cite{Basko,Imbrie,Ehud_rev,Alet_rev, Huse_rev, Abanin_rev} up to a finite temperature. In the MBL phase, the system ceases 
to act as its own bath due to its non-ergodic nature and hence an isolated quantum system in the MBL phase cannot thermalize~\cite{Huse_rev,Abanin_rev}. 
Thus, the lack of ergodicity which is generally identified using the statistics of level spacing ratios~\cite{mehta_random, oganesyan_huse_2007, Alet} 
and the violation of eigenstate thermalization hypothesis~\cite{Deutsch,Srednicki,Rigol}, is among the crucial characteristics of the MBL phase. 
The lack of ergodicity is also reflected in quantum quench studies where the system in the MBL phase shows a strong memory of 
the initial state. This has made the time evolution of the density imbalance a popular tool to analyze the MBL phase 
both theoretically~\cite{Luitz,Yevgeny_rev,Mirlin-imbalance,Mirlin-AA,garg_lr,sierant1,garg_rlr} and experimentally~\cite{dynamics-expts}. 
The MBL phase has also been shown to have local integrals of motion~\cite{Abanin} which are exponentially localized operators 
which commute with each other and the Hamiltonian.

The delocalization to localization transition is also tracked using the statistics of eigenfunctions~\cite{Alet_rev, Serbyn, Santos, Yevgeny-fractal, Mirlin-fractal}, 
scaling of subsystem entanglement entropy~\cite{Alet_rev, Huse2013, Bardarson, Bera, Alet, Sdsarma, Naldessi, garg} and extremal statistics of entanglement eigenvalues, 
as recently proposed~\cite{Kedar}. Since the MBL transition is a dynamical transition that involves many higher excited states, all the analysis of eigenfunction 
statistics or the entanglement entropy is done for the entire many-body spectrum and not only for the ground state.
Dynamical quantities like the return probability, which gives the probability with which a quantum particle comes back to its initial position at a later time~\cite{Luca, Benini, Subroto, garg, garg_SG, garg_lr,sierant2}, time dependent density-density correlation functions ~\cite{Yevgeny_rev, Bera2, Soumya, Yevgeny2020, Yevgeny-AA} and low-frequency conductivity~\cite{Yevgeny_rev, Agarwal, Subroto, Pixley, Subroto2} calculated in the infinite temperature limit have also been useful to identify the delocalized and the MBL phase.

In this work we study the single particle excitations through the analysis of single-particle Green's function calculated for many-body eigenstates in the middle of the spectrum and in the infinite temperature limit. We demonstrate that the delocalization to MBL transition can be tracked using the typical values of the local density of states (LDOS) of single-particle excitations and the single-particle scattering rates.
To the best of our knowledge, single particle excitations have not been explored in the context of MBL though recently there have been works on the evaluation of the propagator in the Fock space by mapping the many-body interacting Hamiltonian of the MBL problem to an effective non-interacting Anderson model~\cite{Tarzia, Logan}. Here, we focus on the single-particle Green's function in real space and show that the typical value of the single-particle LDOS and scattering rate is finite in the delocalized phase where single particle excitations can propagate over all the allowed many-body eigenstates while in the MBL phase the typical values of the single-particle LDOS and the scattering rates are vanishingly small. The probability distribution of the LDOS and the scattering rate in the delocalized  phase is very well approximated by a log-normal distribution while it becomes a very narrow distribution in the MBL phase.
The transition point obtained from this analysis is consistent with the one obtained from the statistics of level spacing ratios. 
We further analyzed the Green's functions for all the many-body eigenstates which carry signatures of the many-body mobility edges that are broadly consistent 
with the location of the mobility edges obtained from the energy resolved statistics of level spacing ratios. 

There have been earlier works on the single-particle LDOS in the ground state of disordered interacting systems (with spinful fermions) using various versions of 
dynamical mean field theory~\cite{dmft} as well as other approaches~\cite{nandini} for $d\ge 2$ Anderson-Hubbard model. Most of these studies 
have focused on the single particle Green's function at $T=0$ in the thermodynamic limit treating interactions perturbatively and disorder exactly. 
They show a transition from single particle Anderson localization to a Mott insulator via an intermediate metallic phase as the interactions 
strength is increased for a fixed disorder strength. In this work, we analyze the single particle Green's functions in the infinite temperature limit of a finite size system using exact diagonalisation. For a one-dimensional model of spinless fermions, treating both interactions and disorder exactly, we calculate the eigenstate Green's function and the self energy for all the many-body eigenstates. Typical values of the LDOS and the scattering rates calculated for the many-body 
eigenstates in the middle of the spectrum as well as ensemble averaged over the entire spectrum carry clear signatures of the MBL transition.
Since many-body states in the middle of the spectrum get localized in the end as the disorder strength increases, our analysis captures salient 
features of the MBL transition, which would be missed if one restricts the analysis only to the ground state.

To be specific, we study the model often used to analyze the MBL transition, namely, the one-dimensional model of spinless fermions in the 
presence of random disorder and nearest neighbor repulsion.  The Hamiltonian of the model studied is
\bea
H=-t\sum_{i}[c^\dagger_ic_{i+1}+h.c.] + \sum_i \epsilon_i n_i + V\sum_i n_in_{i+1}
\label{model}
\eea
with periodic boundary conditions. Here, the onsite energy $\epsilon_i \in [-W/t,W/t]$ is uniformly distributed, 
with $W$ as the disorder strength, $V$ is fixed to $t (=1)$ in the entire analysis and the system is half-filled. 
We solve the model using exact diagonalization, for several system sizes from $L=12$ to $L=20$, to obtain the full 
set of energy eigenvalues $E_n$ (for all system sizes) and eigenfunctions $|\Psi_n\ra$ (till $L=18$).
The model in \Eqn{model}, which can also be mapped to a model of interacting spin-1/2 particles by 
the Jordan-Wigner transformation~\cite{footnote1}, has been extensively studied in the context of MBL and a 
transition from the delocalized phase to the MBL phase is seen as the disorder strength $W$ increases~\cite{Alet}. 
However, to the best of our knowledge, the analysis of the single-particle excitations, their LDOS and the scattering rate for this model in 
the infinite temperature limit to look for signatures of MBL has not been attempted before and this is the main focus of our work.

The Green's function in the $nth$ eigenstate is defined as $ G_{n}(i,j, t) = -i \Theta(t) \langle \Psi_n  \vert \{ c_i(t), c_j^\dg(0)\} \vert \Psi_n \rangle$
where $i,j$ are lattice site indices. In the Lehmann representation, one can write the Fourier transform of $G_n(i,i,t)$ as
\be
G_n(i,i,\omega) = \sum_m \frac{|\la\Psi_m|c^\dagger_i|\Psi_n\ra|^2}{\omega+i\eta-E_m+E_n} +\frac{|\la\Psi_m|c_i|\Psi_n\ra|^2}{\omega+i\eta+E_m-E_n}
\label{Gn}
\ee
Here if $|\Psi_n\ra$ is the $nth$ eigenstate of the Hamiltonian in \Eqn{model} for $N_e$ particles in the chain, states $|\Psi_m\ra$ used in the 
first (second) terms in \Eqn{Gn} are obtained from the diagonalization of $N_e+1$ ($N_e-1$) particle systems. $\eta$ is a positive infinitesimal 
and is set to $10^{-2}$ in our simulations to decide a finite broadening such that sufficient number of eigenstates fall in a bin of width $\eta$.
The local density of states is defined as $\rho_{n}(i, \omega) = \left( - \frac{1}{\pi} \right ) Im \left [ G_{n}(i, i, \omega) \right]$ and the self energy matrix is obtained from generalized Dyson equation $\boldsymbol {\Sigma}_n (\omega)\equiv \mbf{{{{G^{-1}_{0}(\omega) - G^{-1}_{n}(\omega)}}}}$ for the $nth$ eigenstate.
Here, $\mbf{G_0}(\omega)$ is the non-interacting Green's function matrix of the disordered system in \Eqn{model}. The scattering rate is identified 
as ${\Gamma}_n(i, \omega) = - Im \left[\Sigma_n (i,i,\omega)\right]$.

In this work, we will mainly focus on the Green's function calculated for eigenstates in the middle of the many-body spectrum to study the MBL transition. 
This is because many-body density of states of \Eqn{model} is sharply peaked in the middle of the spectrum for sufficiently large system, 
and hence an infinite temperature limit, which basically gives the average over the entire spectrum, will have a dominant contribution from 
states in the middle of the spectrum. At the end, we also present results obtained by averaging over the entire many body spectrum, 
which gives an exact infinite temperature limit of the Green's function and the scattering rates.

\begin{figure*}[t]
\begin{center}
\includegraphics[scale=0.6]{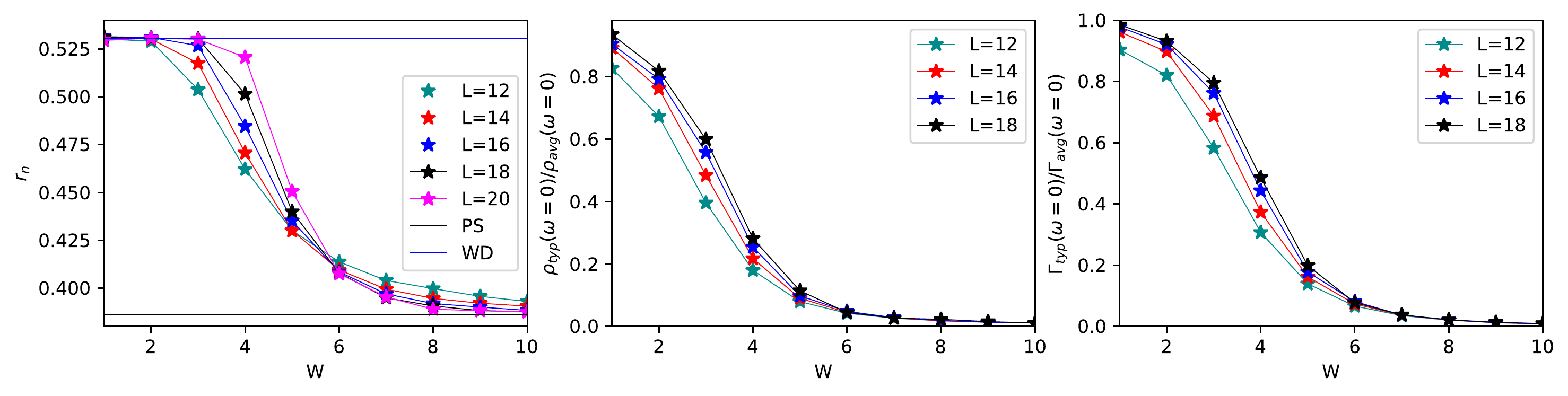}
\caption{{\small{First panel: The disorder averaged level spacing ratio as a function of disorder strength $W$ for various system sizes. Delocalization to MBL transition occurs at $W \sim 6.0t$. Second panel: the ratio of the typical to average local DOS $\rho_{typ}(\omega=0)/\rho_{avg}(\omega=0)$ at $\omega=0$, as a function of the disorder strength. The ratio is of order one for $W \ll W_c$ and increases with $L$. For $W > W_c$, the ratio is vanishingly small and does not show any system size dependence. A similar trend is seen in the ratio of typical to average value of the scattering rate $\Gamma_{typ}(\omega=0)/\Gamma_{avg}(\omega=0)$. All quantities are computed for states in the middle of the eigenspectrum for a rescaled energy bin $E \in [0.495,0.505]$}}}
\label{Fig1}
\end{center}
\end{figure*}

\begin{figure*}[t]
\begin{center}
\includegraphics[scale=0.55]{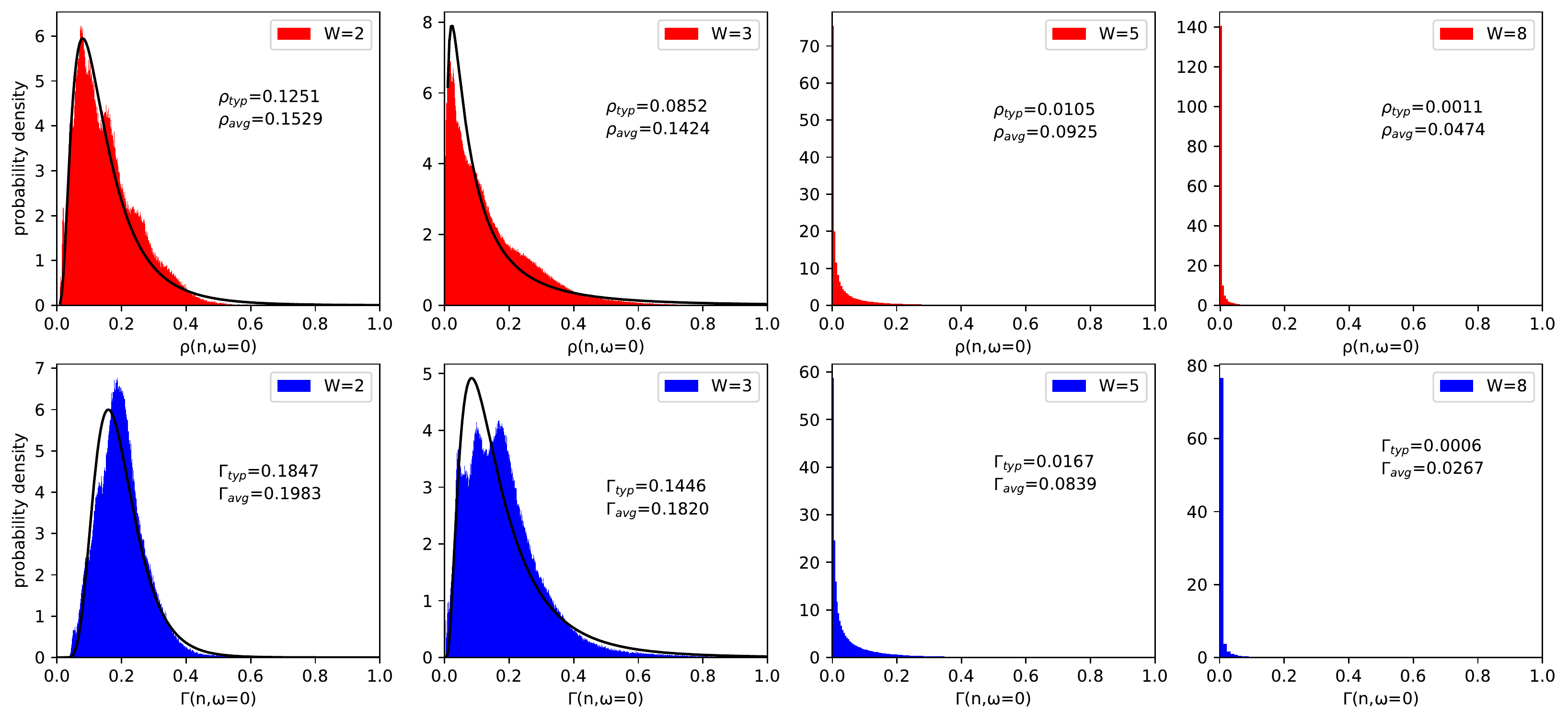}
\end{center}
\caption{{\small{
Probability distribution functions of the LDOS and the scattering rate at $\omega=0$ for disorder strengths $W=2,3,5,8$ and $L=18$. These probability densities have been calculated in the middle of the spectrum for a rescaled energy bin $E \in [0.495,0.505]$.}}}
\label{Fig2}
\end{figure*}
Furthermore, we compare these results with the behavior of a commonly used diagnostic of the transition, namely the statistical behavior of level spacing ratios $r_n = \frac{min(\delta_n, \delta_{n+1})}{max(\delta_{n},\delta_{n+1})}, \textrm{~where,~} \delta_n = E_{n+1} - E_n $.

The typical values for LDOS and scattering rates are obtained by calculating the geometric average over all the lattice sites and independent disorder realizations, 
e.g. the typical value of the LDOS for the n-th eigenstate is given by, $\displaystyle{\rho_{typ}^{}(n, \omega) =  \big[\prod_{C_{\alpha}}\prod_{i=1}^L \rho_i(n, \omega)\big]^{1/CL}} $, 
where $C$ is the number of independent disorder configurations and $C_{\alpha}$ denotes a particular configuration. The definition for the typical scattering rate 
${\Gamma}_{typ}(n, \omega)$ is completely analogous. We obtain the disorder averaged values by averaging over a large number of independent disorder configurations 
details of which are given in the Supplementary Material (SM)~\cite{supplemental}.

\Fig{Fig1} depicts a comparison of these quantities as a function of the disorder strength in the middle of the many-body energy spectrum. 
The transition from ergodic to non-ergodic behavior is clearly seen in the disorder averaged level spacing ratio $r_n$ as the transition from 
Wigner-Dysonian (WD) to Poissonian statistics (PS). For $W < W_c \sim 6.0t$, as expected, the disorder averaged $r_n$ increases with the 
system size approaching the average value for the WD distribution $(\approx 0.5295)$ while for $W> W_c$, $r_n$ decreases as the system 
size increases and approaches the average value for the PS $(\approx 0.3863)$. The second and third panels of \Fig{Fig1} show the most important results of our work, namely,
the ratio of the typical to the average value for the LDOS $\rho_{typ}(\omega=0)/\rho_{avg}(\omega=0)$ and the scattering rate 
${\Gamma}_{typ}(\omega=0)/\Gamma_{avg}(\omega=0)$ calculated at the middle of the many-body spectrum and for $\omega=0$. In the delocalized
phase for very weak disorder, the typical value is of the order of the average value, both for the local DOS and the scattering rate. As the
disorder strength increases, while still being less than $W_c$, the ratio of the typical to the average value increases with the system size.  
In marked contrast to this, in the MBL phase for $W > W_c$, the typical value of the local DOS and the scattering rate becomes much smaller 
than the corresponding average values such that the ratio of typical to average values for both the quantities show a clear approach to 
almost zero without any significant dependence on the system size.

To understand this, let us consider creating a particle-hole pair on top of a many-body eigenstate $|\Psi_n\rangle$. The resulting many-body state can be written as a linear combination of various many-body eigenstates. If $|\Psi_n\rangle$ is a localized eigenstate, the number of eigenstates contributing to the excited state is also of measure zero. Hence, the excitation can not propagate over all the many-body eigenstates allowed by the energy conservation such that $\langle \Psi_m|c^\dagger_i|\Psi_n\rangle$ and $\langle \Psi_m|c_i|\Psi_n\rangle$ vanish. Thus the typical LDOS (as obtained from Eq.~\ref{Gn}) for low-energy single particle excitations is vanishingly small for many-body localized states. On the other hand, if $|\Psi_n\rangle$ is an extended state, the excited state obtained by creating a particle-hole pair on it will get contributions from a significant fraction of many-body eigenstates making the typical LDOS finite in the delocalised phase. Note that the average value of LDOS remains finite even in the MBL phase due to rare region effects which make the distribution broad and asymmetric (as shown below) though the average value also decreases as the disorder strength increases. The quasiparticle relaxation rate is governed by the typical value of the imaginary part of the self energy $\Gamma$ which is of the order of broadening $\eta$ in MBL phase while it is finite in the delocalised phase.

There are some common features in the behavior of the infinite temperature LDOS for the MBL system and the LDOS of the non-interacting Anderson localized systems in higher dimensions~\cite{Thouless,Abu_chakra,Cohen,Ohkawa, Mirlin_AL1, Mirlin_AL2, Janssen,Fehske1, Fehske2, Fehske3, Song, Vollhardt,AL_book}. In both the cases the typical LDOS acts as the order-parameter across the localization transition rather then the average LDOS. This is simply because the physical quantities in a disordered system are broadly distributed and the average values are not characterisitc for the distributions with long tails. But there are crucial differences in the physics of interacting and the corresponding non-interacting system. In non-interacting disordered systems, the LDOS is simply  $\rho_0(\omega)=\sum_n |\Psi^0_n|^2\delta (\omega-E_n)$ where $|\Psi^0_n\rangle$ is the single particle wavefunction. In the strongly localized phase of the Anderson transition, the single particle wavefunction is exponentially suppressed at sites away from the localized site which results in vanishingly small values of the typical LDOS~\cite{Thouless,Abu_chakra,Cohen}. However, in the interacting system, LDOS of the single-particle excitations is a nontrivial function of many-body wavefunctions (See Eq.~\ref{Gn}), which are most naturally localized or extended in the Fock space. Hence, for the non-interacting problem LDOS follows the scaling of the inverse participation ratio~\cite{Brillaux} but the LDOS for the MBL system does not in general follow the scaling of the many-body wavefunctions. For $V=0$, there are no low energy single-particle excitations that can propagate in the Fock space for any strength of disorder in 1-d for the model in Eq.~\ref{model}. The presence of finite interactions helps in creating propagating single-particle excitations first in the middle of the many-body spectrum and then in the entire spectrum as the interaction strength $V/W$ increases.

\begin{figure*}[t]
\begin{center}
\includegraphics[scale=0.6]{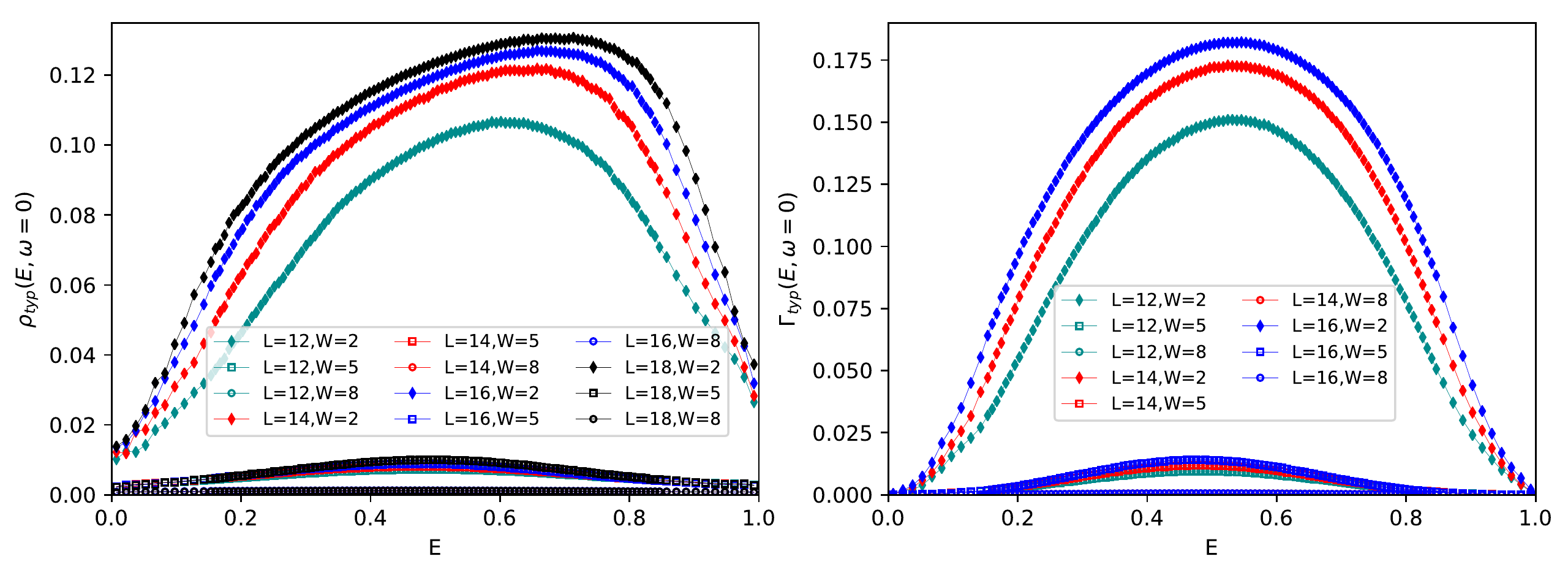}
\end{center}
\caption{{\small{
The typical LDOS $\rho_{typ}(E,\omega=0)$ and typical scattering rate $\Gamma_{typ}(E,\omega=0)$ vs the rescaled eigen-energy $E$ for three disorder strengths and various system sizes. 
Typical value of the scattering rate first vanishes for the eigenstates at the edges of the spectrum and a much larger disorder strength is required to make it vanishingly small for 
the states in the middle of the spectrum. A similar picture is depicted in the left panel which shows the typical LDOS vs $E$, and both of these are consistent with the disorder averaged energy 
resolved level spacing ratios shown in the SM. The topmost curves are for $W=2t$ and lower sets are higher disorder values $5t$ and $8t$ respectively.}}}
\label{Fig3}
\end{figure*}

It is instructive to investigate the complete probability distribution of the LDOS and the scattering rate rather than just looking 
at the typical and the average values. In \Fig{Fig2} we have shown the probability distribution functions of $\rho_{i}=\rho_n(i,\omega=0)$
and the scattering rate $\Gamma_i = \Gamma_n(i, \omega=0)$ for eigenstates with $E_n$ in the middle of the many body spectrum. 
For weak disorder, both the quantities have broad distributions with the arithmetic mean and the typical value being close to the 
most probable value of the distribution. Fits of our numerical data (shown in the figure as solid lines) reveal that the distribution 
functions are close to a log-normal distribution for both the quantities in the delocalized phase.
It is interesting to note that near the localization transition the $2+\epsilon$ dimensional Anderson model is known to have 
log-normal distribution of LDOS \cite{Janssen}. Though for the non-interacting Anderson localization, the log-normal distributions are 
associated with multifractality of critical wavefunctions and the LDOS \cite{Janssen}, for the interacting system even deep inside the delocalized 
phase we find probability distributions to be close to log-normal, where the eigenfunctions are extended in conventional sense.  

As the disorder strength increases such that $W < W_c$, the peak of the distribution shifts towards lower values and 
a long tail develops which is induced due to enhanced rare region effects. As $W$ increases further beyond $W_c$, more weight gets transferred to extremely low values of both 
the LDOS and the scattering rates and the width of the distribution reduces significantly.
Thus in the strong MBL phase the distribution is almost a delta distribution and the typical value vanishes but the 
arithmetic average remains finite. This trend of the probability distribution of the scattering rates is qualitatively 
consistent with the analysis by Basko et. al~\cite{Basko}. Though we have primarily focused on the properties of low energy excitations, that is, $\omega \sim 0$ behavior of the LDOS and the scattering rates, we have also investigated the higher energy single-particle excitations through finite frequency behavior of the typical LDOS $\rho_{typ}(\omega)$ and the typical scattering rate $\Gamma_{typ}(\omega)$ (shown in Fig. 3 in the SM \cite{supplemental}).
More or less the same features as the zero frequency case are seen in the LDOS and the self energy for finite $\omega$ as well.
This is clear from the fact that at all $\omega$ within the band width, $\rho_{typ}(\omega)$ as well as $\Gamma_{typ}(\omega)$ 
decreases as the disorder strength increases.

So far we have presented results for various physical quantities calculated in the middle of the many-body spectrum. 
In order to see whether the LDOS and the scattering rate carry signatures of a transition across the many-body spectrum 
and particularly whether one can identify many-body mobility edges with these quantities, we analyzed the single particle 
Green's functions in the entire many-body spectrum. \Fig{Fig3} shows the typical value of the LDOS $\rho_{typ}(E, \omega=0)$ and scattering rate $\Gamma_{typ}(E,\omega=0)$ vs the rescaled energy $E$. 
As the disorder strength increases, first the typical value of the LDOS and the scattering rate at the edges of the spectrum vanishes and it requires much 
stronger disorder strength to make the quantities at the middle of the spectrum vanishingly small. For $W>W_c$, the typical LDOS as well as the scattering rate is vanishingly small 
over the entire many-body eigenspectrum. This is qualitatively consistent with what is observed in the disorder averaged energy-resolved 
level spacing ratios shown in the SM~\cite{supplemental}. A careful look at \Fig{Fig3} shows that for the localized states the typical value 
of the scattering rate is less then the broadening $\eta$. In contrast to this for the delocalized states in the middle of the spectrum the typical LDOS and the scattering rate 
increases with the system size for weak disorder and $\Gamma \gg \eta$. In terms of the ratio of the typical to average values of the LDOS and the scattering rate, 
shown in the SM \cite{supplemental}, we see that the ratio is of the order one for states in the middle of the spectrum and 
decreases for states on the edges of the spectrum, giving a picture consistent with the mobility edges obtained from level spacing ratios. Finally, we present results obtained by ensemble average over 
the entire eigenspectrum, which is equivalent to an exact infinite temperature calculation. As shown in the SM~\cite{supplemental}, the 
behavior of the typical values of the ensemble averaged LDOS and the scattering rate is completely analogous to the 
corresponding mid-spectrum quantities. 

In summary, we studied the single particle Green's function in the infinite temperature limit across the delocalization to MBL 
transition in a one-dimensional system of spin-less fermions. We demonstrated that single-particle excitations carry clear signatures of MBL to delocalization transition. 
In the delocalized phase, where the single-particle excitations can easily propagate over the entire Fock space, the scattering rate and the LDOS have broad log-normal distributions with a finite most probable value. In contrast, deep inside the localized phase, where it is not possible for a single-particle excitation to propagate over all the many-body eigenstates within the allowed energy window, both the scattering rate and the LDOS have a delta function like distributions peaked around zero. The MBL transition point obtained 
from the analysis of the single-particle Green's functions and scattering rate is close to the one obtained from the level spacing statistics. 
We further showed how many-body mobility edges can be identified from energy resolved typical scattering rates and LDOS.

Both the quantities studied in this work can be measured in experiments directly. 
Given the recent developments in the field of optical lattices, it has become possible to measure single-particle spectral 
functions in ultracold lattices in disordered potentials~\cite{expt_akw}. We hope very much that such experiments can be 
extended also for the MBL systems which will shine light on the LDOS obtainable by integrating spectral functions in momentum 
space, and the scattering rates which determine the width of the spectral functions.

\acknowledgments
A.G. acknowledges Science and Engineering Research Board (SERB) of Department of Science and Technology (DST), India under grant No. CRG/2018/003269 for financial support.
A.J. and V.R.C acknowledge funding from the Department of Atomic Energy, India under the project number 12-R\&D-NIS-5.00-0100.

\end{document}


\renewcommand{\ni}{{\noindent}}
\newcommand{\dprime}{{\prime\prime}}
\newcommand{\be}{\begin{equation}}
\newcommand{\ee}{\end{equation}}
\newcommand{\bea}{\begin{eqnarray}} 
\newcommand{\eea}{\end{eqnarray}}
\newcommand{\la}{\langle}
\newcommand{\ra}{\rangle} 
\newcommand{\dg}{\dagger}
\newcommand\lbs{\left[}
\newcommand\rbs{\right]}
\newcommand\lbr{\left(}
\newcommand\rbr{\right)}
\newcommand\f{\frac}
\newcommand\e{\epsilon}
\newcommand\ua{\uparrow}
\newcommand\da{\downarrow}
\newcommand\mbf{\mathbf}

\newcommand{\Eqn}[1] {Eqn.~(\ref{#1})}
\newcommand{\Fig}[1]{Fig.~\ref{#1}}

\title{Supplementary material for ``Local density of states and scattering rates across the many-body localization transition"}
\author{Atanu Jana$^{1}$, V. Ravi Chandra$^{1}$, Arti Garg$^{2}$} 
\affiliation{$^{1}$ School of Physical Sciences, National Institute of Science Education and Research Bhubaneswar, HBNI, Jatni, Odisha 752050, India}
\affiliation{$^{2}$ Theory Division, Saha Institute of Nuclear Physics, 1/AF Bidhannagar, Kolkata 700 064, India}
\vspace{0.2cm}
\maketitle
\section{Details of Disorder Averaging: Typical LDOS and Scattering rate}
We compute the disorder averaged values of all the quantities defined in the main text. For this, the energy eigenvalues of each disorder realization are rescaled between $[0,1]$ and averages over disorder realizations are carried out for the same rescaled energy bin. 
For level spacing ratios we use averaging over $15000, 10000, 4000, 500$ and $50$ realizations of disorder for $L=12, 14,16,18,20$ respectively. For evaluation of typical values for the middle of the band we use 8000, 1000, 500 and 50 disorder realizations for $L=12,14, 16 \textrm{~and~} 18$ respectively. The typical values for LDOS and scattering rates are obtained by calculating the geometric average over all the lattice sites and independent disorder realizations, e.g. the typical value of the LDOS for the n-th eigenstate is given by, $\displaystyle{\rho_{typ}^{}(n, \omega) =  \big[\prod_{C_{\alpha}}\prod_{i=1}^L \rho_i(n, \omega)\big]^{1/CL}} $, where $C$ is the number of independent disorder configurations and $C_{\alpha}$ denotes a particular configuration. The definition for the typical scattering rate ${\Gamma}_{typ}(n, \omega)$ is completely analogous. For calculations in the middle of the spectrum, we use a small bin $E=0.5\pm0.005$ centered around the rescaled energy $E=0.5$ and the typical values for that bin are evaluated using a geometric average over all the eigenstates within this bin before doing geometric averaging over lattice sites and disorder realizations. For the analysis of LDOS and scattering rates over the full band we use a similar method and divide the whole range $E=\left[0, 1\right]$ into small bins and evaluate the typical values for each bin.
\begin{figure*}[t]
\begin{center}
\includegraphics[scale=0.35]{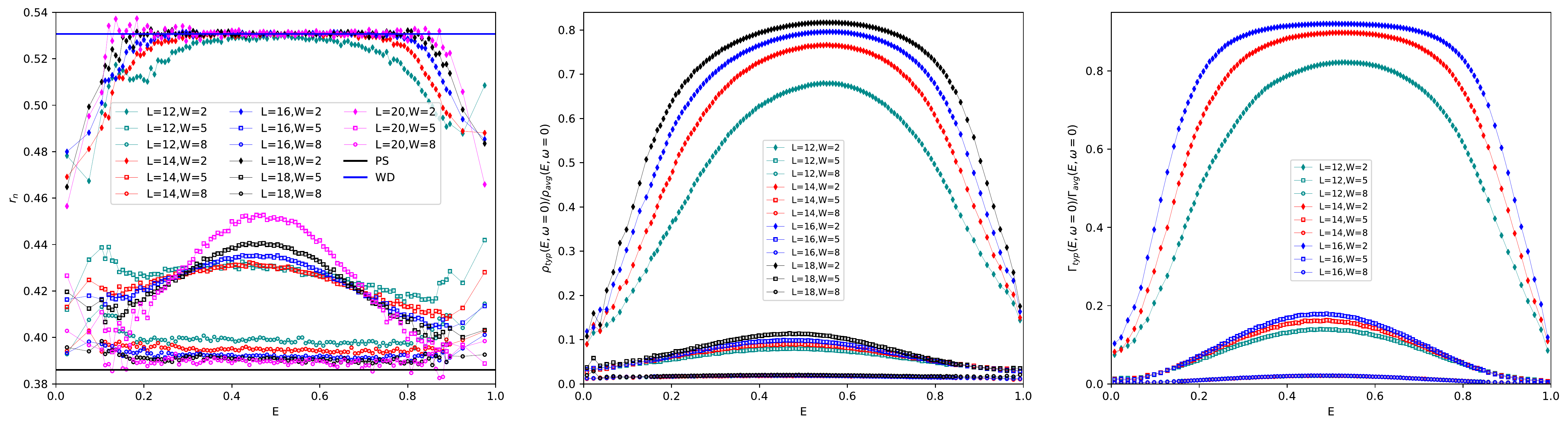}
\caption{{\small{First panel: The disorder averaged level spacing ratio as a function of rescaled energy $E$ for various system sizes and disorder strengths $W=2,5,8$. Middle panel: the ratio of the typical to average local DOS $\rho_{typ}(\omega=0)/\rho_{avg}(\omega=0)$, as a function of $E$. The ratio is of order one for $W \ll W_c$ and increases with $L$. For $W \gg W_c$, the ratio is vanishingly small and does not show any clear system size dependence. Similar trend is seen in the disorder averaged ratio of typical to average value of the scattering rate $\Gamma_{typ}(\omega=0)/\Gamma_{avg}(\omega=0)$, shown in the right most panel.}}}
\label{Fig1_supp}
\end{center}
\end{figure*}

\section{Many body mobility edges}
In the main text in Fig. 3 we analyzed the behavior over the entire energy spectrum of the typical values of the LDOS and the scattering rate. We now compare the behavior of these quantities with that of the level spacing ratio as a function of energy. In \Fig{Fig1_supp}, in the left panel we show the disorder averaged level spacing ratio as a function of rescaled energy for three different disorder strengths and different chain lengths. 
Because of the different statistical limits, Wigner Dyson or Poissonian, for delocalized and localized states respectively we expect a crossing 
point in the energy resolved level spacing plot to be an indication of a many body mobility edge for a given
disorder strength. The second and third panels of \Fig{Fig1_supp} show the ratio of the typical to average values 
of the LDOS and the scattering rates respectively. We see clearly that apart from the general trend of the ratios 
of these quantities becoming small as the disorder strength is increased, their values become nearly independent of the chain 
length in the same region of the eigenspectrum where we find localized states with Poissonian Statistics (PS) in the level spacing plot. 
Thus, the single particle Green's functions also provide evidence of the many body mobility edges that is consistent with the hints given by the statistics of 
level spacing ratios. The level spacing ratios shown in \Fig{Fig1_supp} are averaged over
many more disorder realizations compared to the Green's function quantities (whose computational cost is substantially more than that of the evaluation
of the spectrum). However, the approximate location of the characteristic energy scale associated with the mobility edges, 
is consistent for all the quantities. For all our calculations the largest error in the level spacing ratios 
(the standard error of the mean) is of the order of $10^{-3}$ and most ratios are substantially more accurate. 
\begin{figure*}[t]
\begin{center}
\includegraphics[scale=0.5]{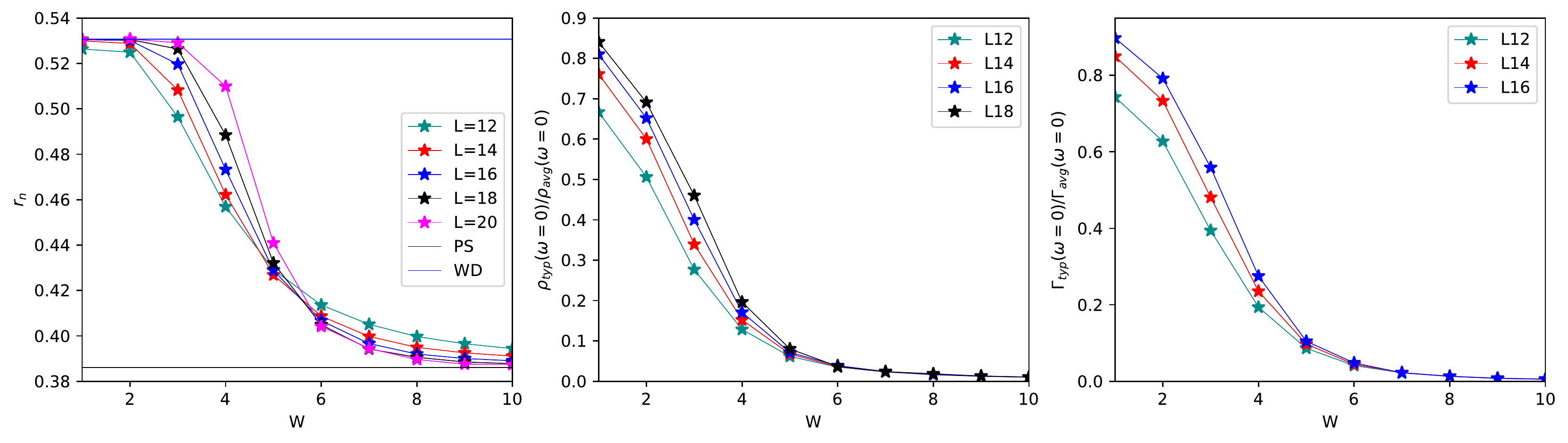}
\end{center}
\caption{{\small{
First Panel: Level spacing ratio averaged over the entire spectrum vs disorder strength. Second and third panels show the ratio of typical to
average values of the LDOS and scattering rates averaged over the entire spectrum vs disorder strength for various system sizes. }}}
\label{Fig3_supp}
\end{figure*}

\section{Infinite temperature ensemble averaged LDOS and scattering rates}
In the main text in Fig.1 we have presented the ratio of the typical to average LDOS and scattering rate, calculated for the mid spectrum eigenstates at $\omega = 0$, as a function of the disorder strength. Here we have shown results obtained by ensemble average over the entire eigenspectrum in the infinite temperature limit. As shown in \Fig{Fig3_supp}, the behavior of the LDOS and the scattering rates calculated from the infinite temperature ensemble average is qualitatively consistent with that obtained from the analysis of mid spectra Green's function. The typical value of the LDOS as well as the typical scattering rate decreases as the disorder strength increases becoming vanishingly small for $W > W_c$ having no significant system size dependence. For comparison we have shown level spacing ratio averaged over the entire spectrum in the first panel of \Fig{Fig3_supp}.

\begin{figure*}[b]
\begin{center}
\includegraphics[scale=0.5]{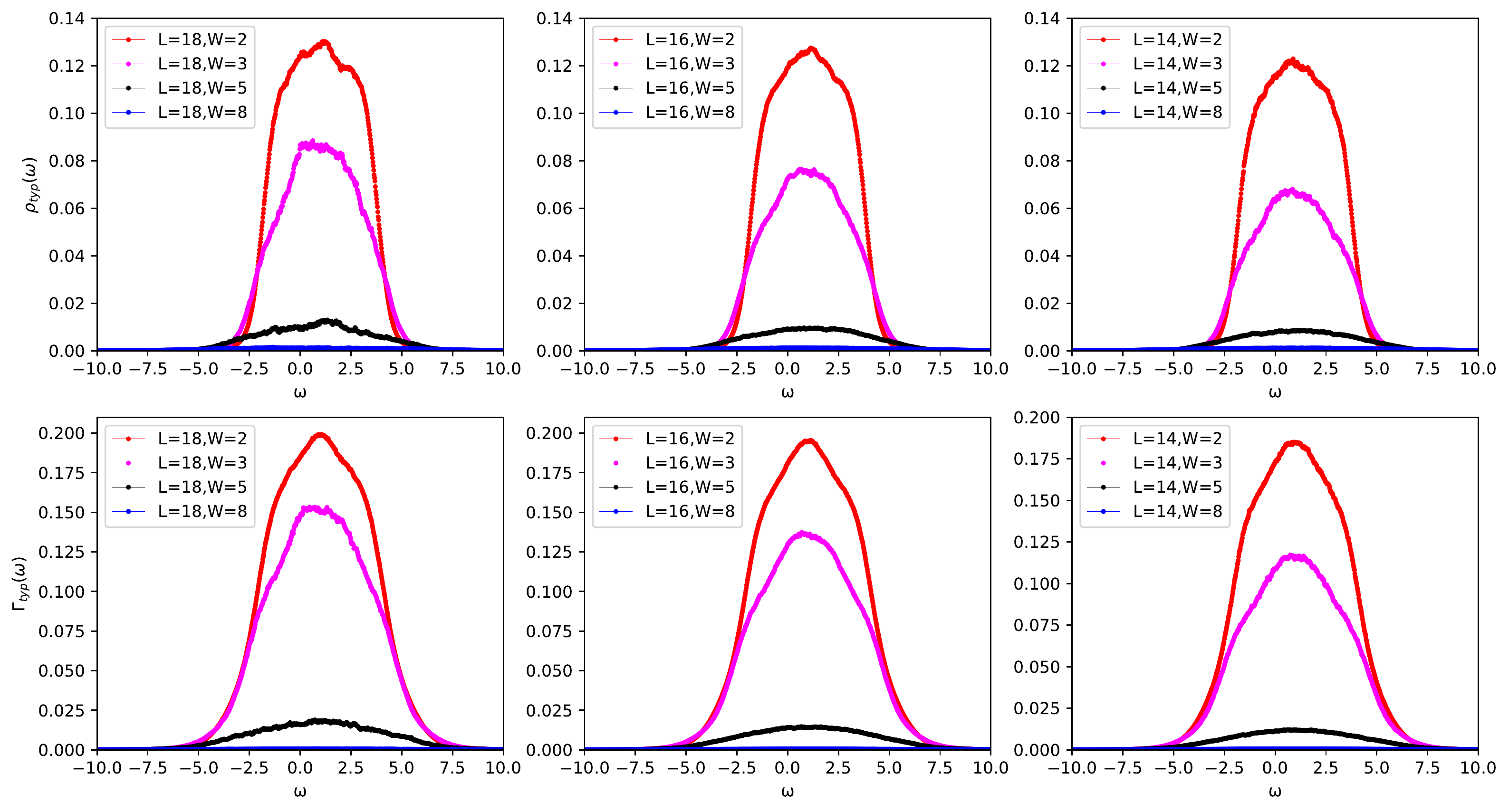}
\end{center}
\caption{{\small{
 $\rho_{typ}(\omega)$ and $\Gamma_{typ}(\omega)$ vs $\omega$ for $L=14, 16, 18$ for various disorder strengths $W=2,3,5,8$. 
The data shown has been obtained using a single state nearest to the rescaled energy in the middle of the many-body spectrum, that is, $E\approx 0.5$ for every disorder configuration.}} }
\label{Fig2_supp}
\end{figure*}
\section{LDOS and scattering rates at finite frequencies}
We have also analyzed the eigenstate Green's functions at a finite frequency which can have a bearing
for example on tunneling experiments with a finite bias voltage. \Fig{Fig2_supp} depicts the behavior
of the finite $\omega$ response for three different chain lengths for different disorder strengths. 
We choose a sufficiently large $\omega$ range to cover all possible contributions from the Lehmann sum for $G_n(i, j, \omega)$.
We see that for the full range of $\omega$ chosen the $\rho_{typ}(\omega)$ and $\Gamma_{typ}(\omega)$ go to vanishingly 
small values for disorder strengths strong enough to localize the system, as in the $\omega=0$ case. The data in \Fig{Fig2_supp}
has been calculated at the middle of the spectrum. Unlike in the $\omega=0$ case, for disorder averaging we use a single state 
nearest to (greater than or equal to) rescaled energy $E=0.5$. Though quantitative details may vary, the general trend of the 
finite frequency disorder averaged $Im[G_n(i, i, \omega)]$ and $Im[{{\Sigma}}_n(i, i, \omega)]$ becoming vanishingly small 
beyond the critical disorder strength will not change if the computationally more expensive evaluation using a 
small energy bin is carried out.